\documentclass[proof]{WileyASNA-v1}

\articletype{Article Type}%

\received{26 October 2024}
\revised{26 October 2024}
\accepted{00 June 2000}

\begin{document}

\title{NSVS 3198272, a low-mass ratio contact binary, revisited\protect}

\author[1,2]{Debski, B.*}

\author[3]{Wysocka, K.}

\author[2]{Skrobacz, W.}

\authormark{Debski, B. \textsc{et al}}

\address[1]{\orgdiv{Astronomical Observatory}, \orgname{Jagiellonian University}, \orgaddress{\state{Orla 171, 30-244 Kraków}, \country{Poland}}}

\address[2]{\orgdiv{Krakowskie Młodzieżowe Towarzystwo Przyjaciół Nauk i Sztuk}, \orgname{H. Jordan Youth Center}, \orgaddress{\state{Krupnicza 38, 31-123 Kraków}, \country{Poland}}}

\address[3]{\orgname{Robert Głębocki Astronomical Observatory}, \orgaddress{\state{Osiek 11/12, 80-180 Gdańsk}, \country{Poland}}}

\corres{*\email{b.debski@oa.uj.edu.pl}}

\abstract{Here we report our study on a low-mass ratio contact binary system NSVS 3198272. We subjected our multi-filter ground-based photometry to the light curve numerical modeling using a modified Wilson-Devinney code. We present three scenarios fitting to the data best: a simple, non-spotted model, a model with a circumpolar spot and a model with a marginal third light. The first two models return similar physical properties which are comparable to the previously reported results on this object. In discussion section we are arguing for advocating the model with a circumpolar spot. At the and we are considering the impact of specific scenarios on the calculated physical parameters of contact binaries. In addition, we report a discovery of a new bright variable star, found by chance as a field star.}




\maketitle


\section{Introduction}\label{sec1}

The UMa-type contact binaries in their standard model \citep{1968ApJ...153..877L} are described as a stable configuration of two main sequence stars that share a common convective envelope. As a direct result of that, both components have virtually the same surface temperature, regardless how small their mass ratio ($q = M_2 / M_1$) is. If the orbital plane of the system is under high inclination ($i > 75^{\circ}$), the observer will record that both primary and secondary eclipses are of similar depths. At the same time, because the components of the contact binary are heavily distorted, the light curve of such a system experiences an extreme ellipsoidal effect. All this results with a constantly varying light curve \citep{1968ApJ...151.1123L}.

Despite the immense complexity of the internal structure of a contact binary, the outer shape of a binary can be very well described with the Roche geometry \citep{kopal}. Such a Roche model can be effectively driven by just the mass ratio, $q$, and the equipotential surface, $\Omega$, that depicts the common outer surface of the binary. The problem of finding these, however, is not always easy due to severe degeneracies between multiple parameters that control the model overall. One of the greatest issues lies in the entanglement between the mass ratio and the inclination of the system. If one has no other means to obtain the mass ratio than the light curve numerical modeling, then the pool of binaries available to study is limited to those experiencing a total eclipse (cf. \citet{2003CoSka..33...38P} and \citet{terrell}). Due to the severe limitation of the possible combinations of light curve's amplitude and the length of the flat-bottom minimum, both the inclination and mass ratio can be established to a high confidence level.

\begin{center}
\begin{table*}[t]%
\caption{Observational characteristics of the variable and the comparison star.\label{tab:obspar}}
\centering
\begin{tabular*}{500pt}{@{\extracolsep\fill}lccD{.}{.}{3}c@{\extracolsep\fill}}
\toprule
Identifier & RA(J2000) & DEC(J2000) & $T$\,[K] & V \\
\midrule
(var) NSVS 3198272     & 20 40 03.6 & +63 59 34.5 & 6618  & 10.70(1)   \\
(comp) TIC 343327204   & 20 39 09.8 & +63 50 46.5 & 6600 & 10.02(2)    \\
(check1) TIC 343409764 & 20 40 14.8 & +63 55 40.3 & 4787 & 11.97(4)    \\
(check2) TIC 343327051 & 20 39 22.7 & +63 59 21.8 & 8398 & 11.94(4)    \\

\bottomrule
\end{tabular*}
\end{table*}
\end{center}

The W UMa-type contact binaries are notorious for their intrinsic variability. Their light curves tend to change shape on a time span of just several weeks. Since the typical orbital period of a contact binary is about $P=0.375\,d$, the time scale of the intrinsic variability can be of just several dozen epochs \citep{debski15,debski22}. A source of such a tentative distortion of the light curve was first hypothesized by \citet{binnendijk} to be a sub-luminous region on the surface of a binary. Not much later \cite{mullan} pointed out that these 'starspots' can be of nature analogous to the spots on the Sun, because of the enhanced depth of the convective zone of a primary (more massive) star in the system. Later on, the spots were proposed as an explanation of the famous O'Connell effect \cite{oconnell}, which is a visible difference in heights between the two consecutive brightness maxima of the W UMa-type light curve. The spots, however, are not easy to model. Adding only one spot to the model introduces additional four degrees of freedom. Because adding two or more spots results a possibly over-driven solution, in this work we will be always assuming that there is only a one, primary 'spotted' region on a surface of a binary. And because cool spots are virtually indistinguishable from hot spots in a simple photometry-based studies, we will be following the model of \cite{mullan} of a cool, magnetic spot that resides on the more massive component.

In this study we are revisiting NSVS 3198272. This totally-eclipsing contact binary was studied in the past by different teams using several different methods for establishing its geometry. This object was considered as a possible contact binary in the Krakow EW-type Light Curve catalog, a precompiled list of object awaiting observational verification\footnote{\url{http://bade.space/ew/}} \citep{debski23}. In our preliminary studies we found the light curve on NSVS 3198272 can be explained using different assumptions with almost equally well fitting models. In this work we are presenting our methods (sec.~\ref{lab:methods}) and our solutions for the physical parameters of both components (sec.~\ref{lab:parameters}). A short discussion on the results and the literature data is presented in sec.~\ref{lab:discussion}.

\section{Observational setup}
NSVS 3198272 was observed as a Target of Opportunity using the Cassegrain telescope (aperture 500 mm, effective focal length 6650 mm) in the Astronomical Observatory of the Jagiellonian University in Krakow on the night of 16/17.10.2019. The photometry was performed in the Bessel B, V, R and I filters with the Apogee Alta U42 camera in the D9 casing. The observations covered the entire orbital period ($P=0.352511737$, \citet{prsa22}) in one run. The near-full Moon enhanced the background sky levels on the second part of the night, which caused the quality of light curve to drop for the last ~30\% of the run.
We have identified one primary comparison star and four control stars. Their and the target's parameters are shown in the Table~\ref{tab:obspar}. Please note the brightness of the object is based on our observations while the comparison stars are taken from the Gaia Synthetic Photometry catalog \citep{gaiaspectra}. The effective temperatures for all stars were taken from the TESS Input Catalog (TIC), v8.2 \citep{tic82}.

\section{Photometry}
The calibration frames were taken before and after observations. We performed a standard Bias-, Dark-, and Flatframe calibration using the C-Munipack v2.1.33 astronomical software\footnote{\url{https://c-munipack.sourceforge.net}} which offers a DAOPHOT-based aperture photometry. We recorded an average size of a star of FWHM\,$\approx 2.4$\,arcsec, which can be classified as a sub-standard quality for the Kraków site.

In this work we identify the primary minimum as the one that coincides with the inferior conjunction, i.e. the primary eclipse occurs when the less massive component (of mass $M_2$) is transiting in front of the more massive primary star (of mass $M_1$). The moment of the primary minimum has been determined with the Kwee \& van Woerden method \citep{kwee} at:
\begin{equation}
    HJD_{Min_I} = 2458773.308817(192)
\end{equation}

\subsection{A new variable star}
We report variability of a bright field star: TIC 343409758. Upon a quick inspection throughout VizieR we found no evidence of this star to have known variability reported earlier. This object showed a constant, flat light curve up to HJD\,=\,2458773.52444, when its brightness started to suddenly drop in a constant rate of 0.31 magnitude per 0.1\,d. Our observations ended when the brightness was still dropping. The overall recorded decrease in brightness lasted 0.153\,d and was of magnitude 0.47. We saw a very small change of the object's color index: $\Delta (B-V) \approx 0.05$. The observed profile of the brightness drop is typical for an eclipse event.

\section{Light Curve Numerical Modeling}\label{lab:methods}
We performed the light curve modeling with a modified Wilson-Devinney code \citep{1971ApJ...166..605W,1985A&A...152...25V}. Our code is equipped with the Price controlled Monte Carlo search method instead of the differential corrections mechanism \citep{1997A&A...324.1010Z}. This software was used earlier in multiple works (e.g. \cite{zola17}, \cite{debski14} or \cite{debski22}). The modeling process assumed a square-root limb darkening law. The coefficients were adopted from \citet{claret11} and \citet{claret13}. The coefficient of the gravitational brightening and the albedo were assumed to comply with the standard accepted values for the convective envelopes of the W UMa-type contact binaries: $\beta=0.08$ (g$\,=0.32$) \citep{1968ApJ...153..877L} and $A=0.5$ \citep{1969AcA....19..245R}. 

\subsection{Establishing the base model}
Our first modeling run assumed a simplistic model with no spots and no third light. To brake the system parameters degeneracy we assumed the temperature of the primary star to be $T_1 = 6618\,K$. With that anchor point we allowed the code to search for the inclination, $i$, temperature of the secondary component, $T_2$, mass ratio, $q$, equipotential surface, $\Omega$, and total light per star in each filter, $L_{f,i}$ (where $f$ denotes the Bessel filter and $i$ stands for the number of a star). We allowed also the model to be phase-shifted with respect to the phased light curves.

The modeling run quickly converged on the model best-fitting to the light curve. Henceforth we will be regarding to this non-spotted model as to the \texttt{Model-1}. The modeling results are stored in the Table~\ref{tab:modelparam}. Please note we report the fill-out factor defined as:

\begin{equation}
    ff = \frac{\Omega_{L_1}-\Omega}{\Omega_{L_1}-\Omega_{L_2}},
\end{equation}

\noindent where $\Omega_{L_1}$ is the Roche equipotential at the inner critical Lagrange surface, $\Omega_{L_2}$ - analogously at the outer critical Lagrange surface, and $\Omega$ is the limiting equipotential of the system (i.e. the surface of the binary). The Roche Model parameters, i.a. the fractional and geometric radii were calculated using the Roche Geometry Calculator available at the website of this project\footnote{\url{http://bade.space/soft/}}.

\begin{figure}[t]
	\centerline{\includegraphics[width=\linewidth]{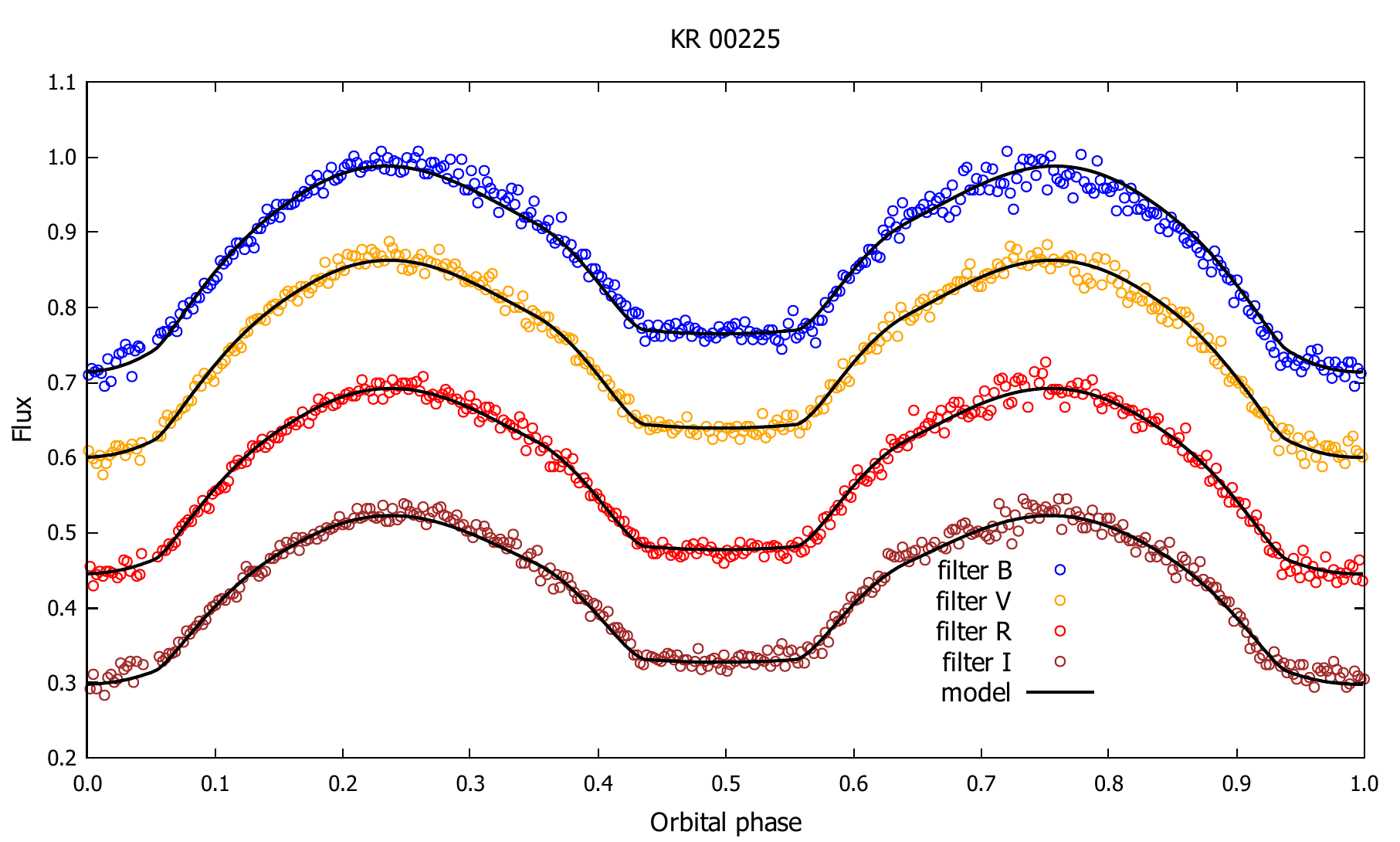}}
	\caption{Observed light curves of NSVS 319827; presented in all four photometric filters, from the top: B, V, R and I. Each dataset is superimposed with its synthetic light curve simulated with \texttt{Model-1} The difference between the synthetic light curves from this and the remaining two models would be indistinguishable in this figure.\label{fig:lc}}
\end{figure}

\subsection{Facilitation of the light curve distortion}
Upon closer inspection, the NSVS 3198272 light curves are slightly asymmetrical, with a barely noticeable O'Connell effect (primary maximum was $\Delta m = 0.02$ higher in the B filter) and both the secondary and primary eclipses are skewed. We conducted a test to see if the modeling software will find a suitable longitude for a cool spot responsible for a light curve distortion of such a type. We allowed the software to search system parameters again, but with a small, cool spot located on the primary component, fixed to its equator. We fixed the temperature of the spot as $T_{spot}=0.75\,T_1$. The radius of the spot remained free for search to complement the fixed latitude and temperature. The resulting best-fitting model returned nearly identical system parameters as in \texttt{Model-1}, with a very small, spot located at the neck between the two stars. This model fitted the light curves exactly as well as the \texttt{Model-1} did. From our experience this means that the modeling run was unable to locate a position of any significant spot. Our reasoning is that the location of a spot on either the back of a star or on the binary' neck can easily be compensated with a minor change in the effective temperature of the secondary star.

To verify if it is possible to facilitate the light curve distortions we decided to choose the spot longitude manually. Besides the light curve shape, the existence of a spot is strongly suggested by the short orbital period of the system \citep{debski22}. The light curve distortion spans noticeably for about 75\% of the orbital phase and affects both minima (skewness) as well as the secondary maximum (O'Connell effect). If this distortion were to be attributed to just one spot, then the sub-luminous region would be of a considerable size. At the same time, the visible O'Connell effect was extremely small and couldn't be caused by a large spot fixed at the equator. The best scenario would be to have a large spot close to the pole of the primary star. We arbitrarily set the radius of the searched spot to $r_{spot}=25^{\circ}$, and its temperature to $T_{spot}=0.85\,T_1$ (following the relation of the star-spot relative temperature from \cite{2005LRSP....2....8B}). The longitude of the spot's center was fixed to $\lambda = 60^{\circ}$. We allowed the software to search for a location of the spot's center between up to 20 degrees from the pole of the primary star (co-latitude $\theta \in (0,20)$).

\begin{figure}[t]
	\centerline{\includegraphics[width=\linewidth]{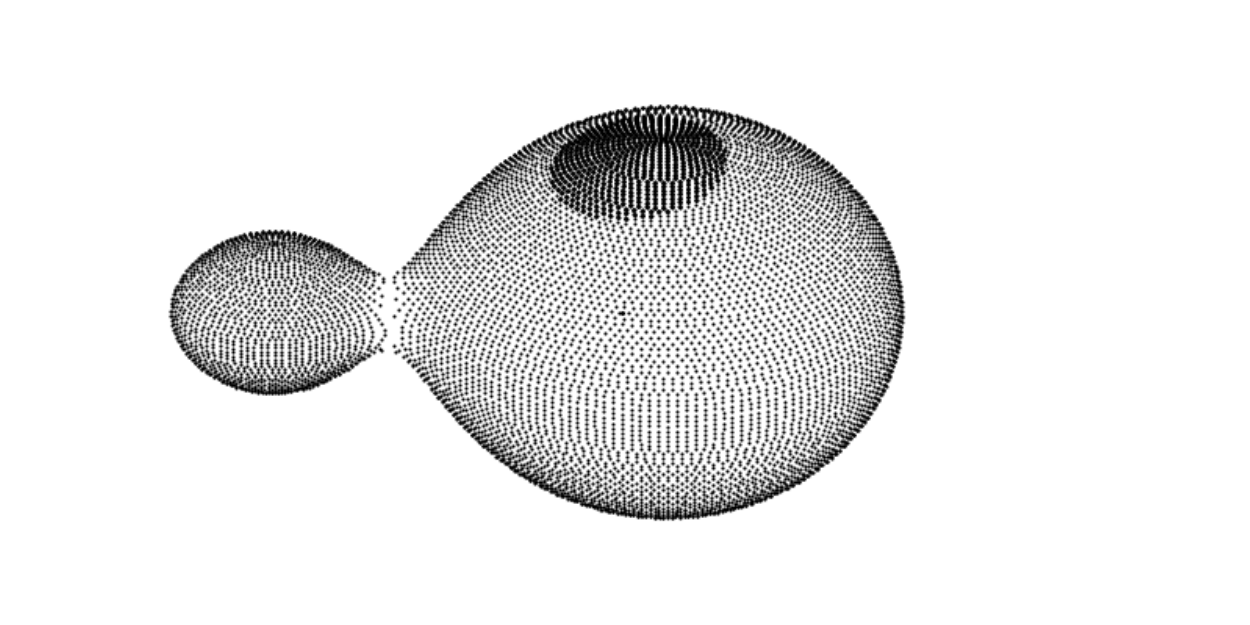}}
	\caption{A visualization of NSVS 3198272 based on the \texttt{Model-2}. Please note the graphical model presented here is under the lowered inclination, $i=70^{\circ}$, only to better present the circumpolar spot to the reader.\label{fig:lc}}
\end{figure}

The best model (hereafter: \texttt{Model-2}) returned a marginally better fit than \texttt{Model-1} (3.4\% lower variance). The model placed the circumpolar spot at a co-latitude $\theta = 12.2(3.5)^{\circ}$ making it effectively a tilted polar cap, visible at any given orbital phase. The presence of a tilted polar spot affected the system parameters in a predictable way: the mass ratio was decreased, the primary star was enlarged and the temperature ratio got lowered. All geometric parameters are shown in Table~\ref{tab:modelparam}.

\begin{center}
\begin{table}[t]%
\centering
\caption{Parameters constituting our three best-fitting models. The \texttt{Model-1} is the simplest, non-spotted, no-third-light model, \texttt{Model-2} takes into account a curcumpolar spot and \texttt{Model-3} fits the light curve with the third light. Each model assumed the primary, more massive component has an effective temperature of $T_1=6618$\,K. \label{tab:modelparam}}%
\tabcolsep=0pt%
\begin{tabular*}{20pc}{@{\extracolsep\fill}llll@{\extracolsep\fill}}
\toprule
 Parameter & \texttt{Model-1} & \texttt{Model-2} & \texttt{Model-3}         \\ 
\midrule
\,$i$ [${}^{\circ}$]   		& $88.9(1.3)$ 	 &  $89.4(1.3)$    & $86.3(2.8)$ 	\\
\,$T_1$ [K] 				& $6618$         &  $6618$         & $6618$          \\ 
\,$T_2$ [K]					& $6435(107) $	 &  $6534(85)$	   & $6463(103)$	    \\  
\,$\Omega$ 					& $2.000(39) $ 	 &  $1.985(32)  $  & $1.972(35) $ 	\\ 
\,$q$, $\frac{m_2}{m_1}$    & $0.125(10) $ 	 &  $0.120(8)   $  & $0.122(13)$   	\\
\,$L_{1,{\rm B}}$ 			& $10.86(8)$   &  $10.87(12)$      & $10.09(55) $ 	\\  
\,$L_{2,{\rm B}}$  			& $1.56(6) $   &  $ 1.63(9) $      & $ 1.52(10) $ 	\\  
\,$L_{1,{\rm V}}$  			& $11.04(7)$   &  $11.07(11)$      & $10.39(59) $ 	\\  
\,$L_{2,{\rm V}}$  			& $1.62(5) $   &  $ 1.67(8) $      & $ 1.60(10) $ 	\\  
\,$L_{1,{\rm R}}$  			& $10.79(7)$   &  $10.82(10) $     & $10.26(57) $ 	\\
\,$L_{2,{\rm R}}$  			& $1.61(4) $   &  $ 1.65(7) $      & $ 1.60(10) $ 	\\
\,$L_{1,{\rm I}}$  			& $9.83(6) $   &  $ 9.86(9) $      & $ 9.28(40) $ 	\\   
\,$L_{2,{\rm I}}$  			& $1.48(4) $   &  $ 1.51(6) $      & $ 1.46(9) $ 	\\ 
\,$l_{3,{\rm B}}$  			& -              &  -              & $ 0.09(5) $ 	\\  
\,$l_{3,{\rm V}}$  			& -              &  -              & $ 0.08(6) $ 	\\  
\,$l_{3,{\rm R}}$  			& -              &  -              & $ 0.07(6) $ 	\\  
\,$l_{3,{\rm I}}$  			& -          	 &  -              & $ 0.06(5) $ 	\\  
\,$\theta$ [${}^{\circ}$] 	& -              &  $12.2(3.5)$	   & -          	    \\   
\,$\lambda$ [${}^{\circ}$] 	& -           	 &  $60.0$    	   & -           	    \\   
\,$r_{spot}$ [${}^{\circ}$]	& -              &  $25.0$    	   & -          	    \\   
\,$T_{spot}$ [$T_1$] 		& -              &  $0.85$         & -                \\ 
\,$ff$  			 		& $42.0(16)\%$   &  $43(14)\%$     & $67(7)\% $    \\
\,$r_{1,{\rm side}}$ 		& $0.590(12) $   &  $0.593(11) $   & $0.601(5) $    \\
\,$r_{2,{\rm side}}$ 		& $0.224(1) $    &  $0.222(1) $    & $0.232(10) $    \\
\,$r_{1,{\rm back}}$ 		& $0.612(12) $   &  $0.615(11) $   & $0.625(3) $    \\
\,$r_{2,{\rm back}}$ 		& $0.270(7) $    &  $0.268(6) $    & $0.292(12) $    \\
\,$r_{1,{\rm pole}}$ 		& $0.529(8) $    &  $0.532(7) $    & $0.536(4) $    \\
\,$r_{2,{\rm pole}}$ 		& $0.214(1) $    &  $0.212(1) $    & $0.220(9) $    \\
\,$r_{1,{\rm geom}}$ 		& $0.576(11) $   &  $0.579(9) $    & $0.586(4) $    \\
\,$r_{2,{\rm geom}}$ 		& $0.235(2) $    &  $0.233(2) $    & $0.246(11) $    \\
\bottomrule
\end{tabular*}
\end{table}
\end{center}

\subsection{Third Light Test}

A considerable number of contact binaries have a third companion. \cite{2006AJ....131.2986P} estimates that the share of contact binaries hosting a third star is between 40\% and 59\%. In agreement with that, \citep{debski22} found that 53\% in their sample of contact binaries exhibited a need for inclusion of a third light in the model due to the extra companion. We have performed a test modeling run to verify, whether the inclusion of a third light would return a different set of system parameters, yet still fitting the data well. For that light curve modeling run we allowed the software to search for the system parameters, a third light, and assumed no spots.

The modeling run converged on a model very similar to the previous two solutions, and the third light was found to be of $l_3=8.3\%\,\pm\,5.1\%$. The goodness of the ft of this model was virtually the same as in \texttt{Model-1}. For the remainder of this work we will regard to this result as to the \texttt{Model-3}. In the next section we will entertain the idea the third light plays a non-negligible role, just to compare, what would be the impact of it on the physical parameters of NSVS 3198272.

\begin{center}
\begin{table}[t]%
\centering
\caption{Physical parameters calculated for each model. First two models share the same absolute amgnitude and total system brightness. The absolute magnitude of the \texttt{Model-3} takes info account the subtracted third light found in the light curve numerical modeling.\label{tab:param}}%
\tabcolsep=0pt%
\begin{tabular*}{20pc}{@{\extracolsep\fill}llll@{\extracolsep\fill}}
\toprule
 Parameter & \texttt{Model-1} & \texttt{Model-2} & \texttt{Model-3}        \\ 
\midrule
$M_V$                   & \multicolumn{2}{@{}c@{}}{3.18(2)} & 3.28(5) \\
$L$ [L${}_{\odot}$]     & \multicolumn{2}{@{}c@{}}{4.57(8)} & 4.17(20) \\
$A$ [R$_{\odot}$]       & 2.62(7)  & 2.60(6)  & 2.45(11)  \\
$M_{tot}$ [M$_{\odot}$] & 1.93(15) & 1.89(14) & 1.59(21)  \\
$M_1$ [M$_{\odot}$]     & 1.72(15) & 1.69(13) & 1.40(20)  \\
$M_2$ [M$_{\odot}$]     & 0.22(3)  & 0.20(3)  & 0.18(4)   \\
\bottomrule
\end{tabular*}
\end{table}
\end{center}

\section{The physical parameters}\label{lab:parameters}
We calculated the physical parameters on NSVS 3198272 for each of three our models. Every time we followed the same procedure, which starts with calculating the system absolute magnitude, $M_V$:

\begin{equation}
M_{\rm V} = m_{\rm V} - A_{\rm V} - 5{\rm log}D + 5 = 3.26(2),
\end{equation}

We take the observed magnitude from our light curve (filter V), the extinction $A_V = 0.355(9)$ from \cite{tic82} and the distance $D=268.5(8)$\,pc from GAIA parallax $\pi=3.724(11)''$ \citep{gaia}. In case of \texttt{Model-3} we subtracted the third light from, hence we report the absolute magnitude in this model to be $M_{\rm V,3}=3.36(5)$.

Next we express the total luminosity of the binary in solar units:
\begin{equation}
\label{eq:lum}
    L_{T}\, [L_{\odot}]=10^{-0.4(M_{\rm V}-4.83)}\,.
\end{equation}

Subsequently, we calculate the orbital separation, or in case of contact binaries, the distance between the centers of mass of the two components. The orbital separation is a scaling factor for the fractional radii established via the numerical modeling. Incorporating that into the Stefan–Boltzmann law we obtain the relation:
\begin{equation}
\label{eq:separ}
    A \,[R_{\odot}] = \sqrt{\frac{L_{T}}{T_1^4r_1^2+T_2^4r_2^2}}.
\end{equation}
Next, we take advantage of the well established orbital period, $P$, and use the Kepler's Third Law to get the total mass of the system expressed in solar masses:
\begin{equation}
\label{eq:mtot}
    M_{tot} \,[M_{\odot}]=\frac{1}{74.53}\frac{A^3}{P^2}.
\end{equation}
Finally, the individual masses can be derived from the total mass of the system and the mass ratio found in the numerical modeling process:
\begin{equation}
M_1 = \frac{M_{tot}}{1+q}.
\end{equation}

All derived by us physical parameters are presented in Table~\ref{tab:param}. 

\section{Discussion and a summary}\label{lab:discussion}

In previous section we presented three models of different level of complexity, and equally well fitting to the observed data. It is a good practice to follow the simplest model if it explains the observations as well as its more elaborated counterparts. We will advocate just that, but we will argue that the model with a spot should not be neglected. Our first argument comes from the small, but measurable O'Connell effect, usually attributable to a star spot. We plan to conduct a follow-up observations of NSVS 3198272 to verify the shape of the light curve five years after our first run. An analysis of the long-base time series from the TESS Mission will also be considered.

Our second argument in favor of the existence of a spot comes from the \texttt{Model-3}. While the magnitude of the third light in this model is comparable to its modeling error in redder part of the spectrum, we see the third light rises above the model uncertainties in bluer filters. An addition of the third light compensates the supposedly lowered amplitude of the observed light curve. The same lowering of the light curve amplitude can be caused by a constant subtraction of the light, taken away by a constantly visible agent. In this case, a circumpolar sub-luminous region.

When compared to the previously reported results for NSVS 3198272, \cite{2019AJ....158..186K} reported $q=0.115(1)$, \cite{latkovic2021} reported $q=0.115$ and \cite{xu} reported $q=0.125$. In the work of \cite{2019AJ....158..186K} the model was not fitting well to the egdes of the flat bottom secondary minimum, which allows us to argue a slight underestimation of the mass ratio. Subsequently, the latter publication derived the system parameters using a Neural Network method. Our models (both the non-spotted and the spotted one) return the same mass ratio, within the bounds of uncertainty.

The last remark must consider the reported absolute masses. We have calculated the physical parameters with a series of assumptions basing on a simplified model and the base Roche geometry. The resulting error propagation is formally correct, but should be taken with a care. In the regime of our results a change in the absolute magnitude of $\Delta M_V = 0.1$ would result in changing the total mass of the system of about $\Delta M_{tot}=0.25$\,M$_{\odot}$. This should be always taken into account when incorporating the final result into the statistical study.

\section*{Acknowledgments}

The data in this work was collected in a project supported by the \fundingAgency{Polish Ministry of Science and Higher Education} Preludium 12 Grant no. \fundingNumber{ 2016/23/N/ST9/01218}.

\subsection*{Conflict of interest}

The authors declare no potential conflict of interests.

\bibliography{Wiley-ASNA}%

\end{document}